\documentclass[reprint,prb,twocolumn,showpacs,superscriptaddress,aps,longbibliography]{revtex4-2}

\usepackage[colorlinks=true,citecolor=blue]{hyperref} 
\usepackage{graphicx}
\usepackage{bm}
\usepackage{amsmath}
\usepackage{amssymb}
\usepackage{comment}
\usepackage{braket}

\DeclareGraphicsExtensions{.png,.jpg,.eps}
\usepackage{xcolor}

\begin{document}

\title{Hamiltonian-learning quantum magnets with non-local impurity tomography}

\begin{abstract}
Impurities in quantum materials have provided successful strategies for learning properties
of complex states, ranging from unconventional superconductors to topological insulators.
In quantum magnetism, inferring the Hamiltonian of an engineered system becomes a
challenging open problem in the presence of complex interactions.
Here we show how a supervised machine-learning technique can be used to infer Hamiltonian parameters from
atomically engineered quantum magnets by inferring fluctuations
of the ground states due to the presence of impurities. We demonstrate our methodology
both with a fermionic model with spin-orbit coupling, as well as with many-body spin models 
with long-range exchange and anisotropic exchange interactions. 
We show that our approach enables performing Hamiltonian extraction in the presence of significant noise, 
providing a strategy to perform Hamiltonian learning with experimental observables in atomic-scale quantum magnets.
Our results establish a strategy to perform Hamiltonian learning by exploiting
the impact of impurities in complex quantum many-body states.
\end{abstract}

\author{Greta Lupi}
\affiliation{Department of Applied Physics, Aalto University, 02150 Espoo, Finland}

\author{Jose L. Lado}
\affiliation{Department of Applied Physics, Aalto University, 02150 Espoo, Finland}

\date{\today}

\maketitle

\section{Introduction}
\label{sec:introduction}

Quantum magnetism provides a highly flexible playground to engineer exotic phenomena~\cite{Savary_2016,Broholm2020}. 
While a variety of natural materials host frustrated and magnetic states, accessing the most unconventional
regimes often requires using artificial platform where Hamiltonians can be engineered~\cite{RevModPhys.88.041002}.
Atomic-scale quantum magnets~\cite{wiley2023, wang2022electronspinqubitplatformassembled, scienceade5050, Eigler_1990, Yang2019, Seifert_2020, Baumann_2015,Yang2021} have opened new avenues for exploring fundamental quantum phenomena,
enabling controllable platform to engineer symmetry breaking~\cite{Loth2012}, quantum criticality~\cite{Toskovic2016},
emergent excitations~\cite{Spinelli2014,PhysRevLett.131.086701}, and topological states~\cite{Zhao2024,Wang2024}. 
Scanning tunneling microscopy (STM) enables creating and probing artificial quantum states with high spatial and energy resolution, 
providing detailed insights into the behavior of quantum magnets~\cite{wiley2023, wang2022electronspinqubitplatformassembled, scienceade5050, Eigler_1990, Yang2019, PhysRevB.109.195415,Mishra2019,Seifert_2020, Baumann_2015,Yang2021}.
Electrically driven paramagnetic resonance with STM has brought quantum control sensing at the atomic scale to remarkable levels.
ESR-STM enabling energy resolution several orders of magnitude below thermal noise~\cite{Baumann_2015,Willke2018},
allowed quantum gate at the atomic level~\cite{Wang2023_gate}, and probing time evolution
of magnetic dynamics with atomic accuracy~\cite{Wang2023_mag}.
Single atom manipulation in quantum magnets thus enables a degree of control
radically different than in bulk materials, motivating the development of
new strategies to tackle open problems in quantum magnetism~\cite{PhysRevApplied.20.024054}.

\begin{figure}[t]
    \centering
        \includegraphics[width=\linewidth]{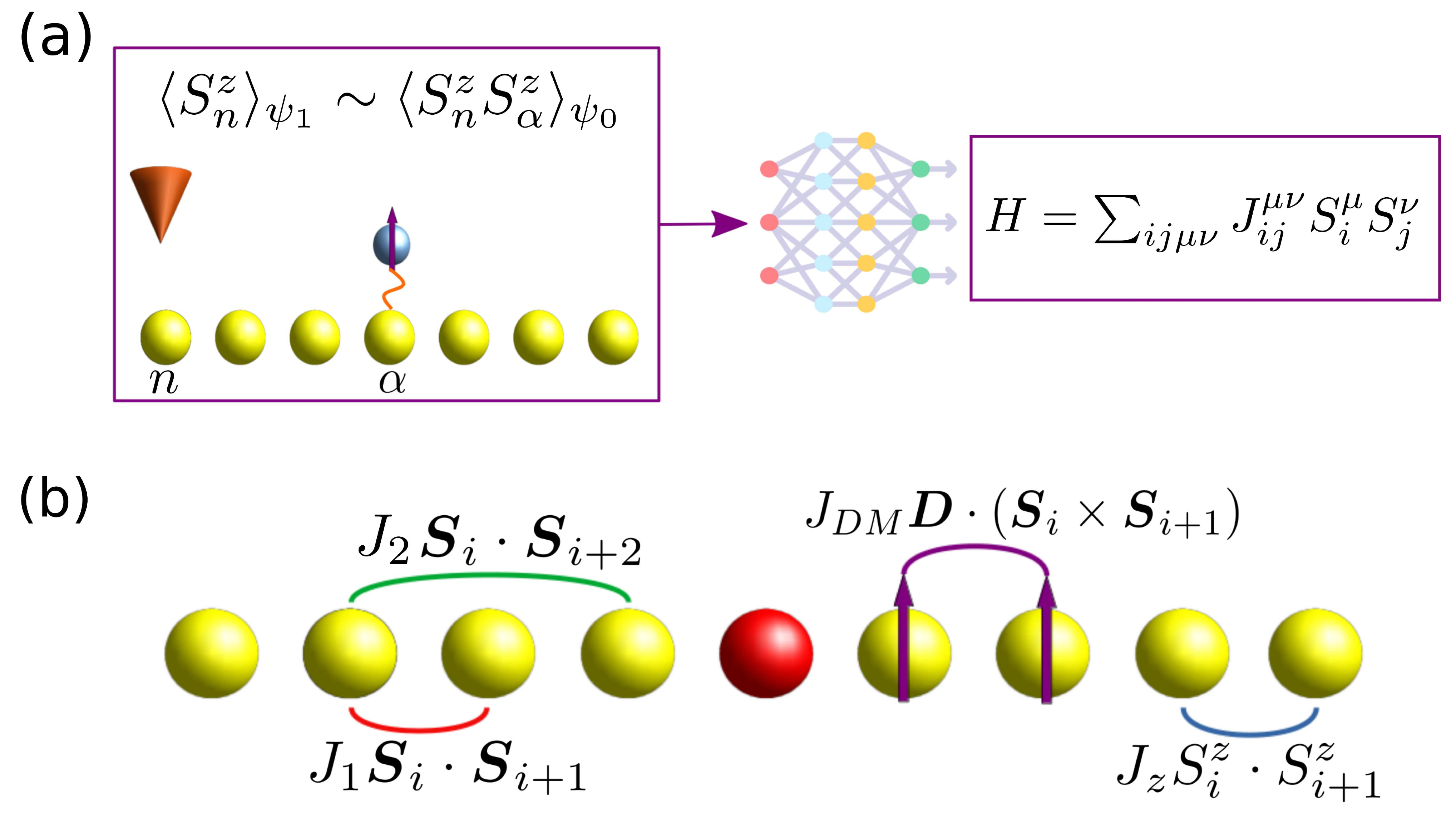}
        \caption{(a) Schematic representation of the Hamiltonian learning strategy 
        that leverages the response of a quantum magnet to perturbations. The inclusion of perturbations
        in a quantum magnet provides access to the non-local correlator,
        that enables learning the original quantum many-body Hamiltonian.
        (b) Shows the specific quantum magnet we will focus on,
        featuring first and second neighbor exchange, anisotropic exchange and
        Dzyaloshinskii-Moriya interaction.}
        \label{fig:pic1}
\end{figure}

Understanding an experimental quantum many-body system relies on
having an accurate description of the many-body interactions. 
Traditional methods of determining Hamiltonian parameters often rely on
fitting experimental data to theoretical models.
Characterizing quantum spin liquids represents a remarkable problem in quantum materials~\cite{Savary_2016,Broholm2020}
due to the intricate nature of their ground state, often requiring new strategies to find their
experimental signatures~\cite{PhysRevResearch.2.033466,PhysRevLett.125.267206,PhysRevB.109.035127,PhysRevB.108.014425,PhysRevResearch.3.033276,PhysRevB.105.195156,2024arXiv240902190T}.
As a result, even in relatively simple quantum magnets, 
determining the Hamiltonian from experimental data
can be remarkably difficult~\cite{RevModPhys.88.041002}.
Machine learning provides a flexible strategy to enable learning of Hamiltonian parameters directly
from physical observables~\cite{PhysRevApplied.20.024054,2024arXiv240506688L,Tranter2024,PhysRevB.110.075402,Khosravian_2024}.
Hamiltonian learning can be done with data generated from theoretical models,
that can be later on directly applied to experimental measurements~\cite{2024arXiv240504596V,PhysRevApplied.20.044081}. 
This strategy exploits the exceptional ability of machine learning methods to
identify non-trivial correlations in observables that enable extracting information about the
underlying ground state and Hamiltonian~\cite{carleo,Wang_2017,2019arXiv191207636E,bairey,10.21468/SciPostPhysCore.6.2.030,Anshu_2021,PhysRevB.109.195125,2025arXiv250109505A,Hangleiter2024,PhysRevResearch.6.L012052,valenti,Koch_2022,Gentile_2021}. 

Recent works have demonstrated that theoretical studies in Hamiltonian learning and spin systems can have direct implications for experimental investigations, particularly in scanning probe microscopy (SPM) and quantum materials characterization~\cite{PhysRevApplied.21.034009,PhysRevApplied.20.024054,PhysRevApplied.18.064074,PhysRevApplied.18.024004,PhysRevApplied.13.054005,PhysRevApplied.20.044081}. 
For instance, theoretical advances in machine learning applied to Hamiltonian extraction have found applications in 
topological magnetic states~\cite{PhysRevApplied.21.034009}, confined quantum magnets~\cite{PhysRevApplied.20.024054}, 
molecular spin qubits~\cite{PhysRevApplied.18.064074}, and even artificial nanoscale superconductors~\cite{PhysRevApplied.20.044081}. 
Our work follows this line of research, showing that machine learning techniques can be exploited to extract Hamiltonian parameters 
directly from local magnetization measurements. This approach not only aligns with recent efforts in theoretical Hamiltonian learning 
but also presents a concrete methodology that can be applied to experimental data.

Here we show how the Hamiltonian of a quantum magnet
can be directly inferred from measurements of the response of the
spin chain to individual impurities.
Our strategy relies on using the response of the system to extract the Hamiltonian
using a machine learning strategy. The inclusion of individual impurities
creates a perturbation in the ground state that directly reflects the underlying Hamiltonian,
a correlation that can be mapped with a supervised neural network (NN).
We show that this methodology enables extracting long-range and anisotropic exchange interactions,
often a challenge to infer in experiments. We show that this methodology can be understood as a generalization
of Hamiltonian extraction in a fermionic model, a procedure that is often performed in STM
via quasiparticle interference.
Our manuscript is organized as follows. In Section~\ref{sec:expt}, we describe the experimental strategy to obtain
the data required for our Hamiltonian extraction. Section~\ref{sec:models} details the many-body spin models we study,
including the fermionic version that provides a minimal exemplification of our methodology. The methodology employed for Hamiltonian parameter extraction using NNs is described in Section~\ref{sec:hamiltonian_learning}. In Section~\ref{sec:noise_models}, we discuss the inclusion of magnetic and scaling noise in the datasets to simulate experimental conditions. In Section~\ref{sec:results}, we present the results of the Hamiltonian inference, 
demonstrating the resilience of our methodology to noise both
in the evaluation as well as during training.
Finally, in Section~\ref{sec:conclusion}, we summarize our conclusions.

\section{Impurity perturbations and non-local response}\label{sec:expt}
Our methodology will rely on using the non-local response of the system to extract the Hamiltonian.
However, STM measures local quantities rather than non-local ones, 
and thus first an strategy must be developed to extract the required non-local information.
Non-local correlators of quantum spin chain
determine the response of the quantum many-body system
to perturbation in the sites,
and thus are reflected when adding additional
impurities that act as exchange sources.
Specifically, additional sites feature strong single-ion anisotropy
as local perturbations, which then enable to measure elsewhere
in the system what the change they created to the ground state.
Experimentally, this was demonstrated with dysprosium single atom magnets~\cite{Singha2021},
enabling atomic-scale magnetic field with stabilities above days.
The Dy impurity creates a local exchange field in the neighboring site, which as
a result modified the expectation value of the magnetization in the rest of the system.
This strategy is exemplified in Fig.~\ref{fig:pic1}a.

The relationship between the response to local perturbations
and non-local correlators can be rationalized via Kubo formalism~\cite{Kubo1957}.
In the following we will consider two Hamiltonians $H_1$ and $H_0$, 
where $H_1$ is a perturbed 
Hamiltonian, and $H_0$ the Hamiltonian
without perturbation.
For the sake of concreteness, we will consider a local perturbation consisting on a local time-dependent
exchange field in site $\alpha$, with the full Hamiltonian taking the form
\begin{equation}
H = \sum_{ij} J_{ij} \boldsymbol{S}_i \cdot \boldsymbol{S}_j + B^z_\alpha (t) S^z_\alpha = H_0 + V(t)
\end{equation}
using time-dependent perturbation theory in the Heisenberg picture,
the change of the expectation value of the magnetization in site $n$
takes the form~\cite{Sakurai2017}

\begin{equation}
\delta \langle \hat S_n (t) \rangle = \int_{-\infty}^{\infty}\chi_{\alpha,n}(t,t') B^z_\alpha (t') dt
\label{eq:kubo1}
\end{equation}

where $\chi_{\alpha,n}(t-t')$ is the non-local time-dependent spin-spin response that can be 
computed with Kubo formalism~\cite{Kubo1957,BruusFlensberg2004,Mahan2000} as

\begin{equation}
\chi_{\alpha,n}(t-t') = -i \theta(t-t') \langle [\hat S^z_n(t) , \hat S^z_\alpha (t')] \rangle
\label{eq:kubo2}
\end{equation}
where $\hat S^z_n(t)$ are the spin operators in the Heisenberg picture,
and $\theta$ the step function.
The previous relationship shows that the change in the expectation value
of the magnetization in site $n$ when a perturbation is added in site $\alpha$
is given by the non-local two-point spin-spin correlator.
This highlights that time-dependent measurements in magnetic spin chains
directly reflect the time-dependent two-point spin correlations
of the quantum magnet ground state.

The relationship between response to a local impurity and the non-local spin-spin
correlator
becomes specially transparent in the limit of a time-independent
perturbation and a strong many-body gap. 
We will consider a system
whose spectra has a typical gap between
ground state and excited states $\Delta$.
We take that the ground state \(\psi_0\) denote the ground state of $H_0$, the system without the external source, 
and \(\psi_1\) the ground state of $H_1 = H_0 + V$, the system with the local
time-independent source, which can
be realized by adding an Ising Dy site to the quantum magnet.
Taking the perturbation $V = B^\alpha_z S^z_\alpha$, the correction in the expectation value of the local magnetization
at site $n$, $\langle S^z_n \rangle_{\psi_1}$ can be expressed with second order perturbation theory as

\begin{equation}
\langle S^z_n \rangle_{\psi_1} \approx B_\alpha^z \sum_{\nu}
f(\epsilon_\nu) \langle \psi_0 | S^z_n | \Psi_\nu \rangle \langle \Psi_\nu| S^z_\alpha | \psi_0 \rangle + \text{h.c.},
\label{eq:corr}
\end{equation}
where we take $\langle S^z_n \rangle_{\psi_0} = 0$ for the unperturbed
expectation value for a quantum magnet, \( |\Psi_\nu\rangle \) represents the excited states of the unperturbed Hamiltonian, $f(\epsilon_\nu) = 1/\epsilon_\nu$ is the prefactor stemming form perturbation theory,
where $\epsilon_\nu$ is the energy difference between the ground state $|\psi_0 \rangle $
and excited states $|\Psi_\nu \rangle$ of $H_0$.
It worth noting that Eq.~\ref{eq:corr} is obtained by taking the spectral representation of the Kubo formula
Eq.~\ref{eq:kubo1} and Eq.~\ref{eq:kubo2} in the static limit.
The contributions from excited states that give rise to a finite
contribution in Eq.~\ref{eq:corr} stem from low energy
states featuring excitations over the ground state that link sites $\alpha$ and $n$. 
Taking that the relevant energy scales of the excitations that correct the original
ground state is $\Delta = \langle \epsilon_\nu \rangle$, we can make the
replacement in Eq.~\ref{eq:corr} $\sum_\nu f(\epsilon_\nu) |\Psi_\nu\rangle \langle \Psi_\nu | \sim \frac{1}{\Delta} \mathcal{I}$.
This allows to rewrite the correction to the magnetization as

\begin{equation}
\langle S^z_n \rangle_{\psi_1} \sim \frac{B^z_\alpha}{\Delta} \langle S^z_n \cdot S^z_\alpha \rangle_{\psi_0}.
\end{equation}
\begin{figure}[t]
    \centering
        \includegraphics[width=\linewidth]{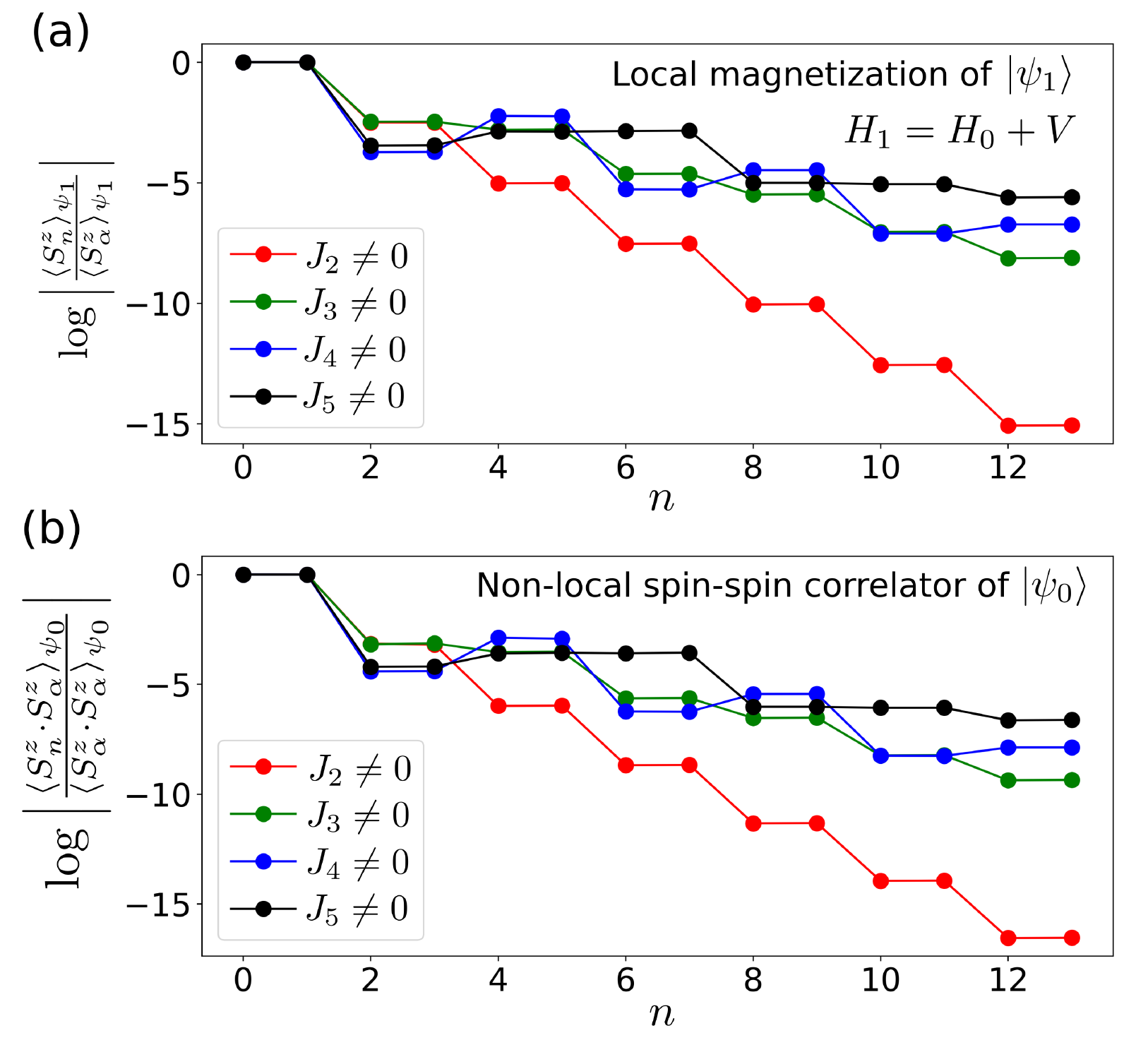}
       \caption{Comparison between (a) the magnetization of a quantum spin chain of Eq.~\ref{eq:longr}
       in the perturbed state \(\psi_1\), with the Ising Dy source coupled at site $\alpha= 0$, and (b) 
       the non-local static correlator of the same spin chain in the unperturbed state $\psi_0$.
       It is observed that both local magnetization and non-local correlator feature analogous
       site dependence, and in particular strongly changing depending on the
       additional long range exchange coupling included.
       We took $\delta = 0.9$, $J_k = 0.2 J_0$ and $B_\alpha^z = 0.01J_0$.}
        \label{fig:pic_mag}
\end{figure}

To factor out the dependence on the strength of the perturbation, the previous result can be normalized by the
perturbed magnetization and unperturbed correlators in site $\alpha$, giving rise to

\begin{equation}
\label{eq:nonloc}
\frac
{\langle S^z_n \rangle_{\psi_1}} 
{\langle S^z_\alpha \rangle_{\psi_1}} 
\approx
\frac{\langle S^z_n \cdot S^z_\alpha \rangle_{\psi_0}}
{\langle S^z_\alpha \cdot S^z_\alpha \rangle_{\psi_0}}
\end{equation}

This relationship allows us to connect the local magnetization \( \langle S^z_n \rangle_{\psi_1} \)
for a gapped static system, which is experimentally measured 
via the differential conductance \( \frac{dI}{dV} \) using spin polarized STM, to the non-local spin correlator \( \langle S^z_n \cdot S^z_{\alpha} \rangle_{\psi_0} \) in the unperturbed system. 
We illustrate the comparison between these quantities in Figure~\ref{fig:pic_mag}.
For the sake of concreteness, we take a Hamiltonian with long-range exchange interaction and dimerization
taking the form
$H = H_0 + V$, where 

\begin{align}
H_0 &= \sum_n J_1(1+\delta (-1)^n) \boldsymbol{S}_n \cdot \boldsymbol{S}_n + \notag \\
&+ \sum_{k=2}^5 \sum_n J_k\boldsymbol{S}_n \cdot \boldsymbol{S}_{n+k}, 
\label{eq:longr}
\end{align}
which features a many-body gap for $\delta\ne 0$ and $J_{k}\ll J_1$ for $k=2,3,4,5$.
Figure~\ref{fig:pic_mag}a shows the magnetization in the different sites upon adding the
local exchange perturbation, whereas Figure~\ref{fig:pic_mag}b shows the non-local spin-spin correlator
in the absence of perturbation. It is clearly observed that both quantities feature very good correspondance
over several orders of magnitude, which allows to infer non-local correlators from measuring local magnetizations
of perturbed quantum magnet. The system features a decay correlation and
magnetization for all the different couplings considered, yet interestingly the type of long range
coupling included dramatically changes the speed of the decay and the oscillations.
In this fashion, even though all those system feature similar many-body gaps,
the magnetization and correlators provide a strong fingerprint of the underlying Hamiltonian.
These unique behavior for the different coupling parameters is what provides
the starting ground to attempt Hamiltonian learning from the observables in Figure~\ref{fig:pic_mag}.

The previous methodology exemplifies how non-local spin correlations can be extracted 
from local STM measurements by engineering local
magnetic sites~\cite{wiley2023, wang2022electronspinqubitplatformassembled, scienceade5050, Reale2024-dk}.
As a result, Hamiltonian learning could be performed either directly by measuring the local response a system
when adding local perturbations on different sites (Fig.~\ref{fig:pic_mag}a), or with the non-local correlator in the unperturbed limit (Fig.~\ref{fig:pic_mag}b).
It is worth noting that this strategy can also be used for single particle systems, just by replacing the
spin operators by fermionic operators and adapting the Kubo formalism accordingly.
In terms of Hamiltonian learning, we will illustrate both single particle and the quantum many-body magnet cases.
For the sake of concreteness, we will perform Hamiltonian learning with local measurements of perturbed systems
for the fermionic case, and with the non-local correlators for the quantum many-body limit,
yet noting that either local perturbed measurements
or non-local unperturbed measurements allow for Hamiltonian learning in both cases.
It is finally worth noting that time-dependent measurements, either local or non-local,
could be used for Hamiltonian learning~\cite{PhysRevLett.124.160502,PhysRevA.104.052431,Lapasar2012,PhysRevA.103.042429,PhysRevResearch.3.023246,2022arXiv220914328W,Harper2020,PhysRevLett.130.200403,PhysRevX.10.011006,Yu2023}.

\section{Models}
\label{sec:models}

In this section, we introduce the two systems where we demonstrate our methodology, 
a fermionic chain and a quantum many-body spin-1/2 chain.
The fermionic chain represents a single-particle model, and
it allows us to demonstrate our methodology as a generalization
of Hamiltonian extraction with quasiparticle interference.
Beyond its application on a conventional metal,
it is worth noting that a single particle Hamiltonian learning
could be performed 
as a quasiparticle description of a
quantum magnet in terms of auxiliary spinons.
The quantum many-body spin chain solved with tensor
network methods represents the more accurate and
challenging case,
and is the case where methodology demonstrates its potential.
In this case, we compute the exact quantum many-body ground and
required expectation values using the full quantum many-body wavefunction
using tensor networks~\cite{dmrgpy,fishman2022itensor}.

\subsection{Fermionic chain with a magnetic impurity}
\label{subsec:fermionic_chain}

We start by considering a single particle model for which we will later perform Hamiltonian learning.
While the single particle case is much simpler than the many-body model, it will allow us to exemplify the
essence of Hamiltonian learning with non-impurity tomography that we will later
exploit in the quantum many-body case. The fermionic Hamiltonian that we consider
takes the form:

\begin{align}
    H &= t_1 \sum_{i=1}^{n-1} c_i^{\dagger} c_{i+1} + t_2 \sum_{i=1}^{n-2} c_i^{\dagger} c_{i+2} + \\ \nonumber
    &+ \mu \sum_{i=1}^n c_i^{\dagger} c_i + i\lambda_R \sum_{i=1}^{n-1} [\hat{z} \times (\boldsymbol{r}_i - \boldsymbol{r}_{i+1})] \cdot \boldsymbol{\sigma} c_{is}^{\dagger} c_{i+1,s'},
\end{align}
where \( t_1 \) and \( t_2 \) are the nearest-neighbor and next-nearest-neighbor hopping parameters, respectively, \( \mu \) is the chemical potential, and \( \lambda_R \) represents the strength of the Rashba SOC~\cite{Bychkov1984OscillatoryEA, rashba, Manchon_2015}. Here, \( c_i^{\dagger} \) and \( c_i \) are the creation and annihilation operators at site \( i \), \( \boldsymbol{r}_i \) is the position vector at site \( i \), \( \boldsymbol{\sigma} \) are the Pauli matrices representing the spin degrees of freedom, and \( s \) and \( s' \) denote the spin indices. The term \( [\hat{z} \times (\boldsymbol{r}_i - \boldsymbol{r}_{i+1})] \cdot \boldsymbol{\sigma} \) captures the spin-orbit interaction arising from the presence of a magnetic impurity.\\
We modeled the fermionic chain with a magnetic impurity positioned at the center, coupled to the chain with a strength of $\delta = 0.4$. 
We consider an infinite system that we solve exactly using a Green's function embedding algorithm.
The observable that will be used as input of the machine learning algorithm
is the local spectral function of the perturbed system, defined as

\begin{equation}
    A(n,\omega) = \sum_s \bra{\Omega(\omega)} c_{n,s} \delta (\omega - H) c_{n,s}^{\dagger} \ket{\Omega(\omega)},
\end{equation}

where $s$ denotes the spin and $\ket{\Omega(\omega)}$ is the ground state of the system at the frequency $\omega$. The previous quantity corresponds to the local density of states in the presence of an impurity. The local impurity
gives rise to scattering of the conduction electrons, that create oscillations of the density of states. Such oscillations are a direct consequence
of the electronic dispersion of the system, and reflect the scattering wavevectors associated to the band structure. This phenomenology 
is directly used in quasiparticle interference to infer Fermi wavevectors from experimental data. In particular,
the Fourier transform of the density of states fluctuations reflects a self-convolution of the electronic structure. We will train a machine algorithm that allows to directly infer the parameters of the Hamiltonian from the spectral function and its Fourier transform.

\subsection{Quantum spin 1/2 model with a spin-1 impurity}
\label{subsec:spin_chain}
The second system that we will consider is a full quantum many-body model,
consisting on a spin-$1/2$ chain with long range exchange and anisotropic interactions.
The Hamiltonian of the spin chain is defined as:
\begin{align}
    H &= J_1 \sum_{i=1}^{n-1} \boldsymbol{S}_i \cdot \boldsymbol{S}_{i+1} + J_2 \sum_{i=1}^{n-2} \boldsymbol{S}_i \cdot \boldsymbol{S}_{i+2} +\\ \nonumber
    &+J_z \sum_{i=1}^{n} S^z_i \cdot S^z_{i+1} + J_{DM} \sum_{i=1}^{n-1} \boldsymbol{D} \cdot (\boldsymbol{S}_i \times \boldsymbol{S}_{i+1})
    \\ \nonumber
    &+ B^z \sum_{i=1}^n S_i^z + B^x \sum_{i=1}^n S_i^x,
\end{align}

where \( J_1 \) and \( J_2 \) represent the nearest-neighbor and next-nearest-neighbor exchange interactions, \( J_z \) is the anisotropic exchange interaction, \( J_{DM} \) denotes the Dzyaloshinskii-Moriya (DM) interaction~\cite{dmi,DZYALOSHINSKY1958241} with \(\boldsymbol{D}\) as the DM vector (which we take with module equal to 1), and $ B^z $ and $B^x$ is an external tunable 
magnetic field that acts along the $z$ and $x$-direction, perpendicular and parallel
to the plane in which the chain lies.
Our Hamiltonian learning strategy requires two different types of atoms, the ones forming the quantum magnet, 
and the Ising magnetic field source. In order to amplify the effect of the perturbation, it is however fitting to consider we include
a $S=1$ site in the middle, denoted as the amplifier site in Fig.~\ref{fig:pic1}c. 
This site leads to a finite local spin susceptible to yield a large response in the 
presence of the proximity to the Ising Dy source.
While extracting the response of the system does not require
having the $S=1$ site,
we will include in the following to demonstrate that the
strategy works in this experimentally relevant case.
The external magnetic field $B^z$ in the model enables
controlling the ground state of the system~\cite{delcastillo2023probingspinfractionalizationesrstm},
and will provide the parameter enabling Hamiltonian inference. 
In the absence of an external magnetic field, the Hamiltonian features
time-reversal symmetry, but broken rotational symmetry due to the anisotropic exchange terms.
It is worth considering two limiting cases in which the non-local correlators would
feature dramatically different behavior.
In the limiting case where only $J_1$ is non-zero, the ground state 
features quantum disorder and realizes a pristine $S=1/2$ Heisenberg model
featuring gapless spinons that give rise to non-local
correlations featuring a power-law decay.
In stark contrast, if only $J_z$ is non-zero, the Hamiltonian realizes an Ising model, that features a ground state with stagger
magnetization, with leads to an oscillating $\langle S_n^z S_\alpha^z \rangle$ correlator.
By including a spin-1 impurity, we introduce localized magnetic interactions that perturb the system, offering insights into how impurities affect quantum correlations and the overall ground state. 

The object that we will use for Hamiltonian learning is the field-dependent non-local correlator, defined as
\begin{equation}
    \chi(n,B_z) = \bra{\Omega ({B_z})} S^z_n S^z_{\alpha} \ket{\Omega ({B_z})},
\end{equation}

where \(\ket{\Omega ({B_z})}\) is the ground state of the system at field $B_z$, \( S^z_i \) is the z-component of the spin operator at site \( n \), and \( S^z_{\alpha} \) is the z-component of the spin operator at the impurity site $\alpha$.
This observable is related to the response of the system to local Ising impurities, as described in Sec.~\ref{sec:expt}. In the gapped limit, it is directly given by the magnetization in the presence of a perturbation, as shown in Eq.~\ref{eq:nonloc}. Figure~\ref{fig:pic_mag} presents results with long-range interactions, demonstrating that the relationship between non-local correlators and the system's response to a perturbation holds in these generic models. 

For Hamiltonian learning, we focus on the experimentally relevant scenario where $J_1$, $J_2$, $J_{DM}$, and $J_z$ are present, corresponding to the determination of four non-trivial parameters of the spin Hamiltonian. 

We compute the static correlator with respect to the impurity placed at the center of the chain, considering systems of length $L=21$, where the central site corresponds to a $S=1$ spin.

\section{Hamiltonian Learning with Neural Networks}\label{sec:hamiltonian_learning}

In this section, we present the methodologies used in our study to explore Hamiltonian learning using machine learning techniques.

Our objective is to develop an algorithm capable of mapping the field-dependent non-local correlators to the Hamiltonian of the system~\cite{PhysRevB.106.104435,PhysRevE.105.L062101,2024arXiv241210502T}. While the inverse step can be performed systematically with a quantum many-body solver, extracting the Hamiltonian from observables requires an alternative strategy. We use a machine learning methodology based on supervised learning for this purpose, for which a training set with Hamiltonians and non-local correlators is required. In Hamiltonian learning, the targets are specifically the parameters of the Hamiltonian that define the interactions within the system.

The Hamiltonian we study follows the general form:

\begin{equation}
    H = \sum_n \Lambda_n \mathcal{H}_n
\end{equation}

where $\mathcal{H}_n$ represents the interaction terms with their corresponding parameters $\Lambda_n$. This formulation allows for flexibility in adding or removing interactions based on the specific physical system under consideration.

The target parameters $\Lambda_n$ are initially sampled uniformly within a specific range, chosen to ensure that the system remains within a stable phase and does not cross phase transitions to a symmetry broken regime. When selecting the dataset, it is essential to ensure that the measured spin correlations provide sufficient information for reliable parameter inference.

The choice of the Hamiltonian model primarily depends on the specific atom hosting the magnetic moment, the substrate used, and the spacing between atoms.  
As a concrete example, Ti atoms in MgO~\cite{Willke2018,Yang2019,PhysRevLett.119.227206,Wang2024} realize a model with finite $J_1$, a contribution from $J_2$, and very small contributions from $J_Z$ and $J_{\text{DM}}$. Both $J_Z$ and $J_{\text{DM}}$ stem from spin-orbit coupling effects, which are relatively weak in Ti and MgO. In contrast, replacing Ti with Ho~\cite{Natterer2017} would significantly enhance the contributions of $J_Z$ and $J_{\text{DM}}$, due to the stronger spin-orbit coupling. More generally, for spins on a surface, the sign and magnitude of $J_1$ and $J_2$ can vary depending on the specific atomic arrangement~\cite{PhysRevLett.111.127203}. This flexibility allows for tuning regimes where $J_Z$ and $J_{\text{DM}}$ are comparable to, or even stronger than, $J_1$ and $J_2$, by an appropriate choice of atom, substrate, and site arrangements~\cite{RevModPhys.91.041001}.  

In our case, we adopt a widely used Hamiltonian for spin systems, incorporating exchange interactions $J_1, J_2, J_Z, J_{\text{DM}}$, which are relevant in many quantum magnetic materials, as described in Section~\ref{subsec:spin_chain}. A key advantage of our approach is its adaptability: the framework can be applied to various experimental setups by modifying the Hamiltonian terms to match the relevant physical interactions. To ensure experimental applicability, the training dataset must be carefully selected so that it covers a broad range of realistic parameters. Our model accounts for diverse experimental scenarios by incorporating different choices of atoms, substrates, and geometric arrangements, ensuring robustness across various conditions.  

We start with the one-dimensional electron gas interacting sites with an impurity (Fig.~\ref{fig:pic1}b). The presence of scatterers allows us to probe the response of the system to perturbations, leading to frequency and spatial features that the NN leverages to extract the Hamiltonian. We consider first- and second-neighbor interactions between the sites and, most importantly, the spin-orbit coupling (SOC). SOC terms present significant challenges for NN learning due to their intrinsic complexity. These terms involve non-linear and anisotropic interactions between spin and orbital angular momentum, increasing the dimensionality of the problem. Additionally, SOC effects are highly dependent on the system's symmetry, making generalization difficult for the NN. The introduction of SOC also leads to complex electron correlation effects, which are not easily captured by NNs.

After a first training of the NN with perfect theoretical data, we performed a re-training with noisy data incorporating magnetic noise. This approach ensures that the trained models can effectively handle and make accurate predictions even when confronted with experimental data featuring magnetic noise, a crucial requirement for their application in practical quantum systems.

The details of the neural network training for both fermionic chain and spin chain can be found in Appendix~\ref{sec:neural_network}.

\subsection{Fidelity Evaluation}\label{sec:fidelity}

To assess the robustness of the NN training against noise, we define the fidelity using the Pearson correlation coefficient~\cite{Khosravian_2024, LIU2022106426, rupp}:

\begin{equation}
    \mathcal{F}_{\Lambda_n} = \frac{\langle \Lambda_n^{\text{pred}} \Lambda_n^{\text{true}} \rangle - \langle \Lambda_n^{\text{pred}} \rangle \langle \Lambda_n^{\text{true}} \rangle}{\sqrt{\text{var}(\Lambda_n^{\text{pred}})} \sqrt{\text{var}(\Lambda_n^{\text{true}})}},
\end{equation}

where \( X_{\text{pred}} \) represents the predicted values, \( X_{\text{true}} \) represents the true values, and \(\text{var}(\cdot)\) denotes the variance. The fidelity quantifies the correlation between the predicted and true values, thereby evaluating the NN’s performance under noisy conditions. The fidelity, \(\mathcal{F}\), is defined on the interval \([0, 1]\), where \(\mathcal{F} = 1\) indicates perfect prediction accuracy, implying \(\Lambda_{\text{pred}} = \Lambda_{\text{true}}\), while \(\mathcal{F} = 0\) corresponds to no predictive accuracy.

\section{Noise models}\label{sec:noise_models}

A crucial aspect of Hamiltonian learning is its robustness in scenarios where the observables feature noise. 
In the following we will consider two different sources of noise. The first source consist on magnetic noise,
meaning a fluctuating background magnetic field that affects all the local magnetization measured.
This first noise source adds a typical random value to each correlator, and may flip its sign in certain cases.
Experimentally, this type of noise contribution may arise from residual fluctuating magnetic fields with a time scale slower than the measurement time~\cite{PhysRevB.75.085310,PhysRevB.91.155301,PhysRevB.101.075403}.
A second source can be an offset
created by the
magnetic field of the tip used in scanning tunneling
microscopy during the measurement of the local magnetization of the sites~\cite{PhysRevResearch.3.043185,PhysRevResearch.2.013032,Seifert2020,PhysRevLett.122.227203}.
From a theoretical perspective, this type of noise expectation values
can also emerge in many-body calculations in case the ground states are not computed with enough accuracy, which
in the case of matrix product states would arise from using a not sufficiently large bond dimension~\cite{orus2019tensor}.
The second type of noise is denoted as scaling noise, which it does not necessarily change the
sign of the magnetization measured,
but rather impacts its magnitude.
This type of noise would arise from uncertainties on
the value of the magnetization when using
a second atom as detector~\cite{Choi2017,Natterer2017,Esat2024}, 
allowing to measure relative changes in the magnetization can be observed,
but the estimation of the absolute value may show fluctuations.
From the perspective of the non-local correlators, the two noise models are included as follows.
\begin{figure}[t]
    \centering
        \includegraphics[width=\linewidth]{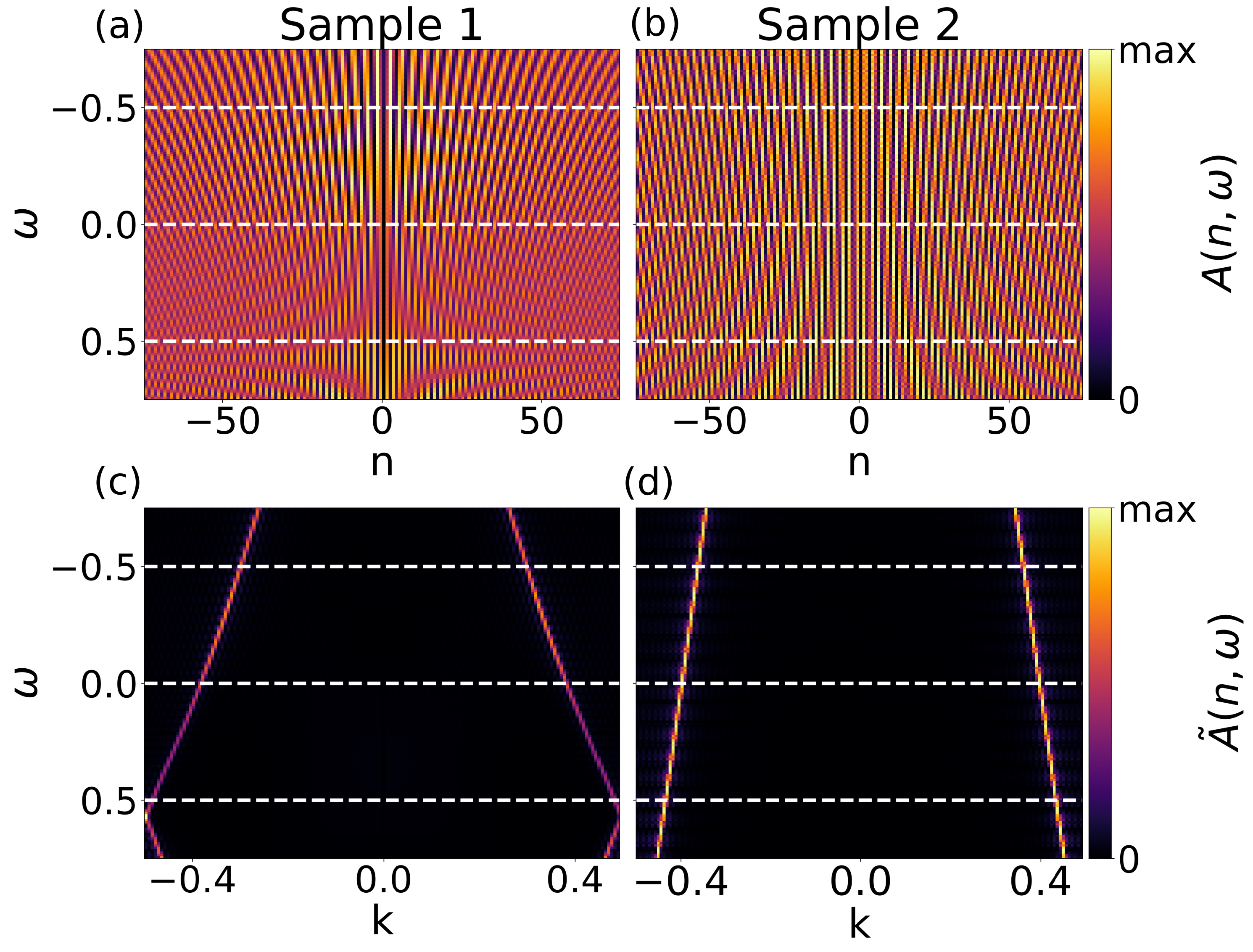}
        \caption{Evolution of the spectral function of two different
        infinite electronic
        system (a,c) and (b,d) with a magnetic impurity.
        Panel (a,b) shows the spectral function in real space $A(n,\omega)$ and
        panels (c,d) its Fourier transform $\Tilde{A}(k,\omega)$,
        as a function of the frequency $\omega$ and, respectively, the position in real-space and momentum-space.
        The different features that appear in each pair (a,c) and (b,d)
        enables extracting the original Hamiltonian.}
        \label{fig:pic2}
\end{figure}
Magnetic noise is included by adding a background contribution to the measured moments as

\begin{equation}
\langle S^z_\alpha S^z_n \rangle_{\text{MagNoise}} = \langle S^z_\alpha S^z_n \rangle + \mathcal{N}(\Delta_B^2),
\label{eq:noisemag}
\end{equation}
where $\langle S^z_\alpha S^z_n \rangle$ represents the original data for the \(i\)-th instance and \(\mathcal{N}(0, \sigma^2)\) 
is the Gaussian noise with zero average and standard deviation $\Delta_B^2$. This type of noise specially impacts long-range correlations,
and in particular limits the accuracy of the potentially extracted correlation length of the quantum many-body state.
In contrast, short range correlations will be much less affected than long-range ones due to its substantially larger value.

The scaling noise takes the form 

\begin{equation}
\langle S^z_\alpha S^z_n \rangle_{\text{ScaNoise}} = \langle S^z_\alpha S^z_n \rangle \left(1 + W \left(n_s\right)\right),
\label{eq:scalenoise}
\end{equation}

where \(n_s\) is a random variable uniformly distributed between $-0.5$ and $0.5$. This noise specially impacts the quantitative values of short range correlations. In contrast, extracted correlation length of the quantum many-body state, that depends on the decay of long range correlations, are relatively robust to this perturbation.

We include this two noise sources at two different levels in our methodology. The magnetic noise 
of Eq.~\ref{eq:noisemag} is included directly in the input data used in retraining the model, and would account both
for experimental residual magnetic noise, and computational inaccuracies of a quantum many-body solver.
The scaling measurement noise of Eq.~\ref{eq:scalenoise} is included only when testing the robustness of the fully trained algorithm,
and would account for uncertainties in the measured magnetic moment due fluctuation current onsets.
In the case of the single particle model, 
noise is included with analogous functional forms, where
the role of magnetic noise is replaced by electronic noise measured in the
spectroscopy~\cite{Ge2019}, and scaling noise stems from potential fluctuations in the
setpoint current~\cite{Hackley2014}.

\begin{figure}[t]
    \centering
        \includegraphics[width=\linewidth]{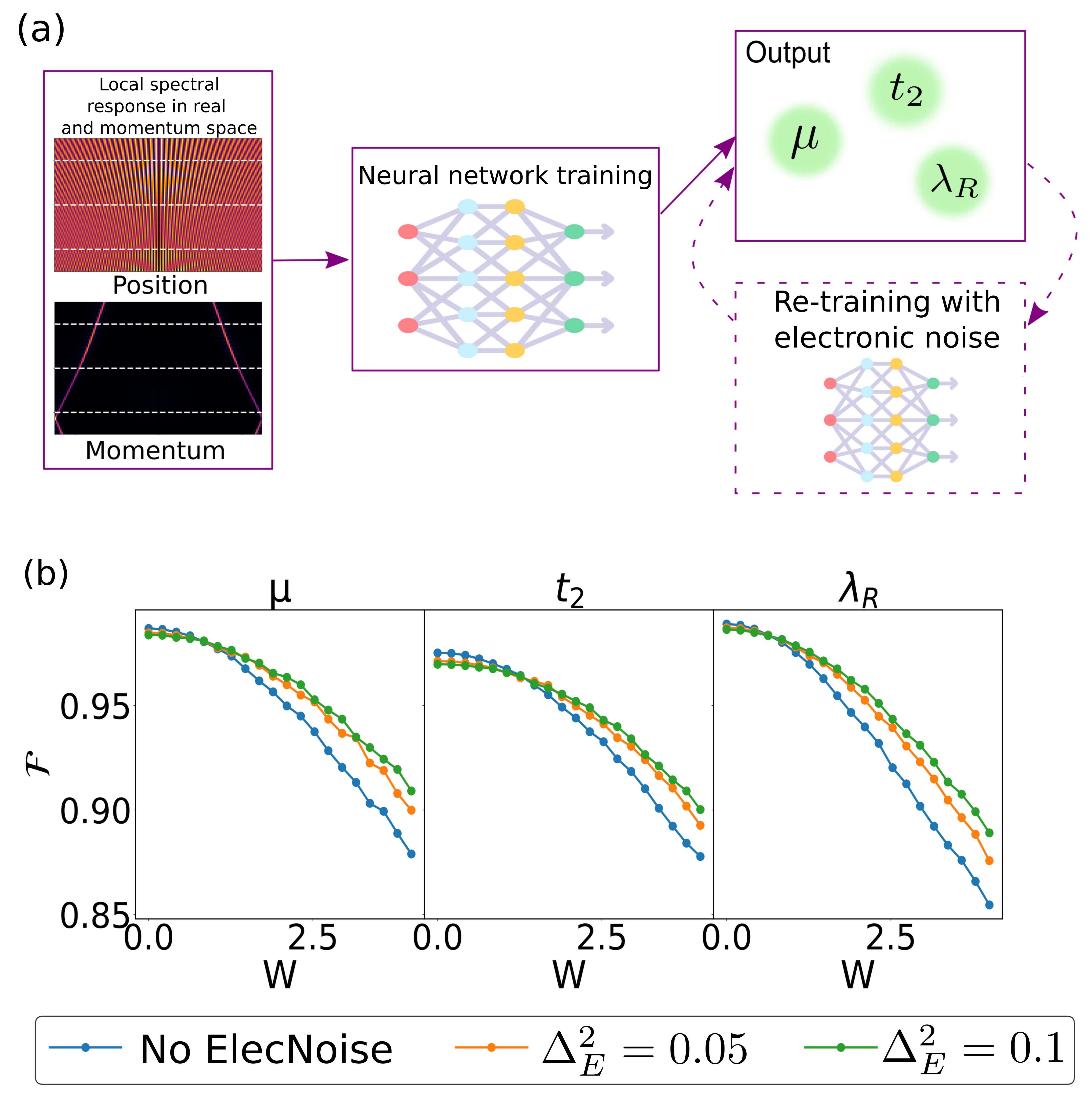}
        \caption{(a) Workflow for the fermionic model. Panel (b) shows the fidelity of the fermionic chain model parameters $\mu$, $t_2$, and $\lambda_R$ as a function of the scaling measurement noise $W$. Each plot compares the fidelity for data without magnetic
        electronic (\(\Delta_E^2 = 0\)) and with electronic noise levels \(\Delta_E^2 = 0.05\) and \(\Delta_E^2 = 0.1\).}
        \label{fig:pic4}
\end{figure}
\begin{figure*}[t]
    \centering
        \includegraphics[width=\linewidth]{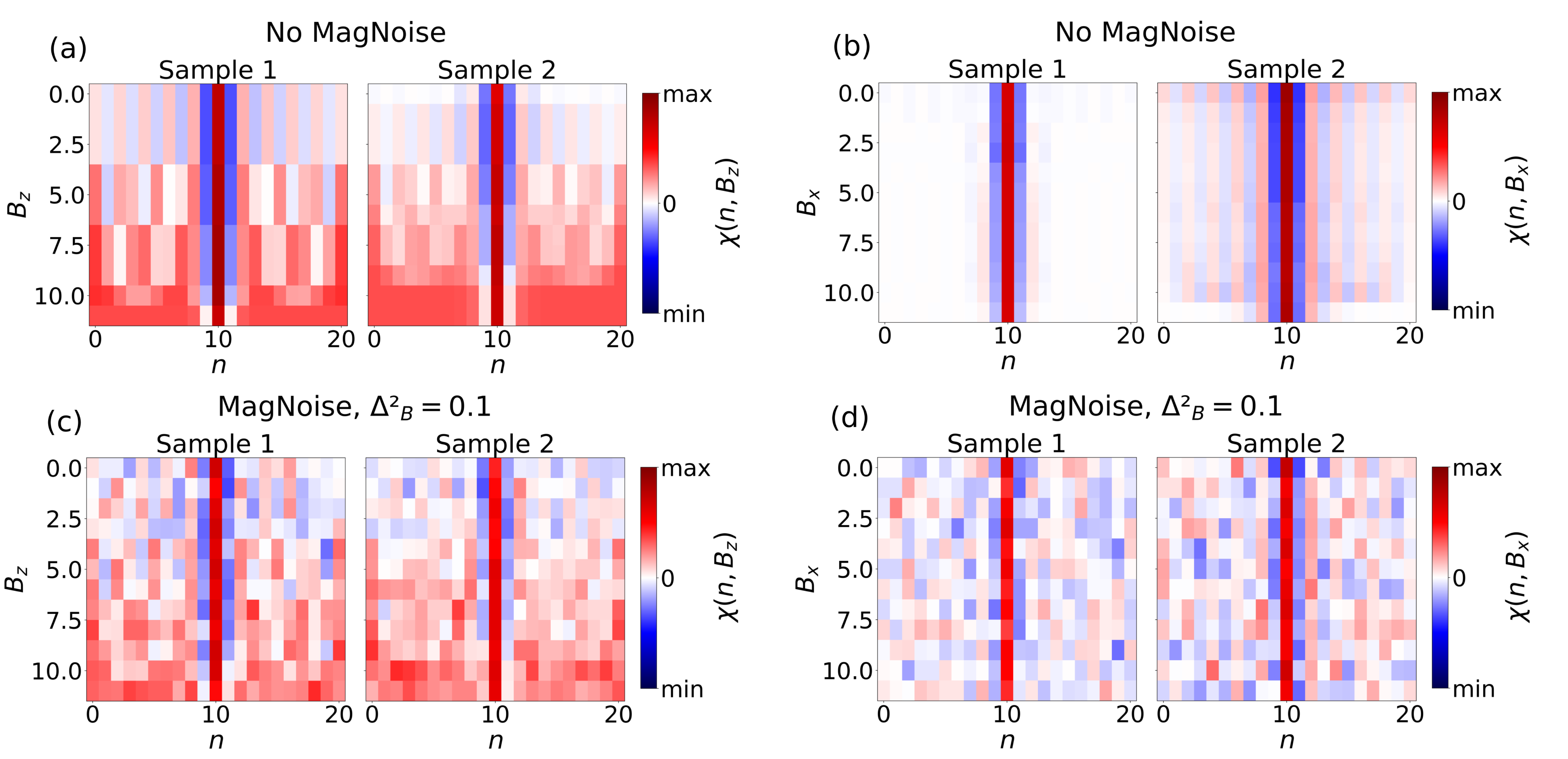}
    \caption{Evolution of the static spin correlator $\chi$ as a function of the external magnetic field \(B^z\) (a,c)
    and the external magnetic field $B^x$ (b,d).
    Panels (a,b) show the pristine spin correlators, whereas
    panels (c,d) show the noisy correlators with magnetic noise at \(\Delta_B^2 = 0.1\).
    The evolution of the correlator with the field acts as
    fingerprint of the  many-body system, 
    which the machine learning algorithm leverages to extract the Hamiltonian. The $n$-axis represents the site index of the spins in the chain.}
        \label{fig:pic3}
\end{figure*}
\section{Noisy Hamiltonian learning}
\label{sec:results}
In this section, we investigate the impact of noise on Hamiltonian learning.
While computational calculations do not feature the same sources of noise as 
experimental ones, the solution of quantum many-body systems
may feature numerical noise to the the approximate nature of the many-body methods.
This is specially relevant in the case of tensor-network states, where the
bond dimension of the variational algorithm determines the accuracy of the ground state.
By examining how noise affects the accuracy and stability of Hamiltonian reconstruction, we can better understand the limitations and potential optimizations of machine learning methods in quantum systems.

\subsection{Single-particle system}
\label{sec:ferms_results}

The spectral function provides information about the electronic structure of the system at different energy levels. By training the NN with the spectral function data, both in real space and its Fourier transform, we can extract essential features and parameters of the Hamiltonian. Fig.~\ref{fig:pic2} shows how the spectral function and its Fourier transform evolve with different biases, in real and momentum space. This problem is fundamentally an image recognition task, where the NN is trained to recognize patterns in the spectral function and Fourier spectral function plots.
The Fourier spectral function enables augmenting the input data,
which is specially useful to extract the subtlety of the Rashba SOC effects. The Rashba coupling introduces a spin-dependent interaction that is often very subtle and difficult to detect directly from the spectral in real space. In fact, Rashba effect can create small shifts and splittings in the electronic structure that might not be easily discernible in real-space DOS plots~\cite{rashba}. By transforming the spectral function data into momentum space, these subtle effects become easier to extract for the machine learning algorithm. This is because the SOC influences the relative distances between peaks in the Fourier transform, making these interactions more detectable.
By training the machine learning algorithm with these simulations, we obtain an algorithm capable
of inferring the Hamiltonian parameters from the local observables.

Fig.~\ref{fig:pic4}b shows the fidelity of the extracted Hamiltonian parameters $\mu$, $t_2$, and $\lambda_R$ as a function of the noise level $w$. The plots compare the fidelity for perfect data, $\Delta_E^2 =0.05$, and $\Delta_E^2 = 0.1$ electric noise levels. As expected, the fidelity shows a gradual decline with increasing magnetic noise. For $\mu$, the model demonstrates high robustness, maintaining a fidelity close to 0.9 even with electric noise $\Delta_E^2 = 0.1$ at scaling noise of $W = 2$. The fidelity for $t_2$ also declines with increasing $W$ but remains quite high overall, with a noticeable drop beyond $W = 3$. The fidelity for $\lambda_R$ is similarly robust, showing a gradual decrease with higher noise levels and maintaining a relatively
high fidelity even at with electronic noise $\Delta_E^2 = 0.1$.

\subsection{Hamiltonian learning a many-body quantum spin model}
\label{sec:spins_results}

We now move to consider the spin quantum many-body Hamiltonian and extract its parameters from the non-local correlators.
In the single particle case, the change in chemical potential allow extracting information of the 
non-local correlators for a ground-state-dependent state.
For the quantum spin model a chemical potential is however not present, and its role is replaced by a external magnetic
field. The tunability of the magnetic field $B$ in the Hamiltonian
allows us to use the static spin correlator as training data for the NN,
and its change with the magnetic field provides the non-trivial information
required to perform Hamiltonian learning. 
By varying the magnetic field and examining the spin correlators across different sites within the chain, we can generate detailed spatial and magnetic field-dependent data,
either by sweeping over the magnetic field in the z-direction (Fig.~\ref{fig:pic3}ac)
or the x-direction (Fig.~\ref{fig:pic3}bd). 
In Fig.~\ref{fig:pic3}a, we show the evolution of the static correlators for different samples, illustrating how the presence of the central
spin-1 impurity affects the magnetic excitations across the chain with and without the presence of magnetic noise. 
Essentially, these images act like fingerprints for specific Hamiltonian parameters, allowing the NN to learn how changes in the magnetic field and spin configurations relate to the underlying Hamiltonian.

Both directions of the magnetic field reflect a non-trivial evolution of the group state, and
for our purpose any of those will enable us to perform Hamiltonian learning.
In the configuration with $B_x$, the static spin correlator shows distinct correlation peaks centered around the middle impurity site, which are more symmetric and localized in the noiseless data. When noise is introduced, the clarity of these features decreases but remains discernible. Compared to the $B_z$ configuration, the $B_x$ field produces different correlation patterns, reflecting the anisotropic nature of the system. Both configurations demonstrate that, although noise impacts the sharpness of the data, key features near the impurity site are robust enough to be used for further analysis, such as training NNs to infer Hamiltonian parameters.

\begin{figure}[t]
    \centering
        \includegraphics[width=\linewidth]{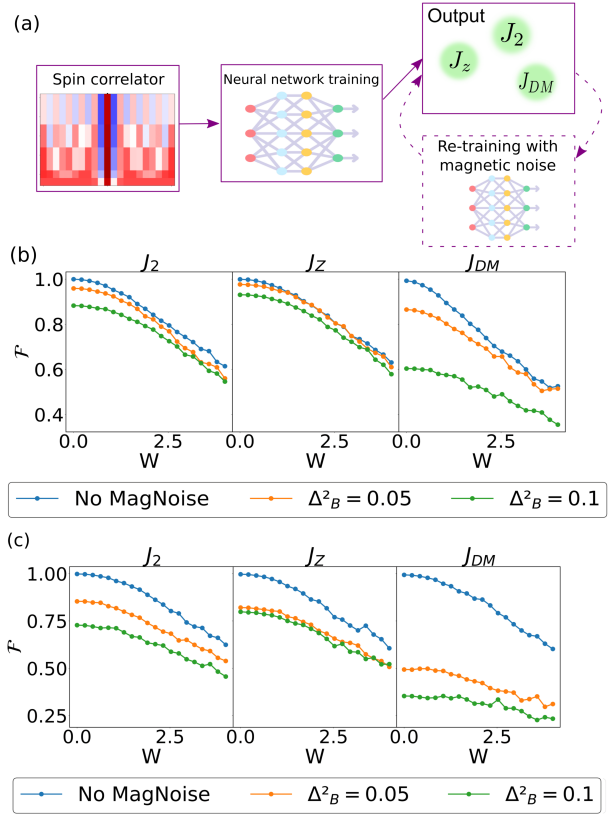}
        \caption{(a) Hamiltonian learning algorithm of the quantum many-body spin model. Panel (b) shows the fidelity of the spin chain parameters $J_2$, $J_z$, and $J_{DM}$ as a function of the scaling measurement noise $W$. Panel (b)
        shows the Hamiltonian learning for a magnetic field in the $z$ direction, and
        panel (c) for a magnetic field applied in the $x$ direction.
        Each plot compares the fidelity for data without magnetic noise
        (\(\Delta_B^2 = 0\)) and with magnetic noise levels \(\Delta_B^2 = 0.05\) and \(\Delta_B^2 = 0.1\).}
        \label{fig:pic5}
\end{figure}

In Fig.~\ref{fig:pic5}b, we present the fidelity of the spin chain model parameters \( J_2 \), \( J_z \), and \( J_{DM} \) as a function of the noise level \( W \) when the magnetic field \( B \) is applied along the \( z \)-direction. This analysis includes three scenarios: perfect data, $\Delta_B^2 = 0.05$, and $\Delta_B^2 = 0.1$ magnetic noise level. As observed with the fermionic chain, fidelity decreases with increasing noise. The parameters \( J_2 \) and \( J_z \) exhibit robustness, maintaining high fidelity even for $\Delta_B^2 = 0.1$ up to \( W = 2 \). Conversely, \( J_{DM} \) is more sensitive, with a sharp fidelity drop at higher noise levels, highlighting its stronger susceptibility to perturbations. In general, magnetic noise further reduces fidelity across all parameters, confirming its substantial impact on learning accuracy.  

Figure~\ref{fig:pic5}c shows results for \( B \) applied along the \( x \)-direction. The overall trend remains similar, with fidelity decreasing as noise increases. However, the parameters display slightly different sensitivities compared to the \( B_z \) case. This is likely due to the reduced variation in the dataset when \( B \) is aligned along the \( x \)-axis, which may limit the model's ability to extract relevant features under noisy conditions.  

The effectiveness of Hamiltonian learning across different interaction regimes is further supported by our results, even in the presence of frustration and gapless spectra. The $J_1$-$J_2$ model, known to exhibit spin-liquid behavior in the antiferromagnetic regime~\cite{PhysRevB.89.241104, PhysRevB.93.144411}, poses challenges due to the potential absence of long-range order. However, our methodology remains reliable in these conditions, thanks to the inclusion of different frustration levels in the training data. This ensures that the model generalizes well across parameter regions, maintaining high fidelity despite minor variations in accuracy.  

A crucial factor affecting the robustness of Hamiltonian learning to noise is the presence of a many-body gap. In gapped systems, low-energy excitations are suppressed, leading to short-range correlations that naturally filter out high-energy noise. This effect enhances the stability of inferred parameters, reducing the impact of fluctuations. In contrast, gapless systems exhibit long-wavelength modes, which introduce extended correlations and amplify noise effects. Consequently, parameter reconstruction may become less accurate, especially when the noise amplitude approaches or exceeds the system’s intrinsic energy scales. Nevertheless, our numerical results indicate that these gap-dependent noise effects do not significantly degrade the overall performance of our machine learning approach, which remains effective across the entire parameter space.  

Finally, we note that more advanced machine learning techniques, such as GANs~\cite{goodfellow2014generativeadversarialnetworks, Koch_2022} and diffusion models~\cite{sohldickstein2015deepunsupervisedlearningusing, NEURIPS2020_4c5bcfec}, could further enhance the efficiency of Hamiltonian learning.

\section{Conclusion}
\label{sec:conclusion}
Here we presented a machine learning methodology to infer Hamiltonian parameters 
that leverages the response of a quantum magnet to local impurity perturbations.
Our strategy exploits the non-local changes created by localized impurities in a quantum many-body wavefunction,
and in particular the dependence of those changes with respect to an externally tunable magnetic field.
Our methodology allows extracting both long-range exchange couplings and anisotropic exchange coupling of quantum magnets,
effects that often represent a remarkable challenge with conventional methodologies.
Our approach offers a powerful method for extracting complex, non-local information from quantum systems, which is often challenging to access through traditional techniques. 
We showed that this methodology
can be carried out in the presence of different noise sources,
showing that this approach is robust even in the presence
of realistic perturbations.
By adapting our methodology to a single
particle problem, we showed that this technique 
enables learning single-particle dispersion, and can be reinterpreted as a machine-learning
inspired quasiparticle interference method. Within the single-particle limit, we showed that this technique allows us to extract spin-orbit coupling terms of the electronic structure from a non-polarized observable, an often challenging effect to extract with conventional quasiparticle interference.

Our method showcases the powerful synergy between machine learning and quantum many-body methods, 
in particular, showing how machine learning techniques trained in computational
data can be used to learn Hamiltonians from experimental observables.
Our results establish a proof of concept to 
apply this machine-learning methodology to more complex quantum many-body systems, 
including higher-dimensional quantum magnets
and interacting fermionic systems. 

\section*{Data Availability}

The datasets and trained models  are publicly available at Ref.~\cite{gretacode}.

\section*{Acknowledgments}

We acknowledge financial support from
the Research Council of Finland Projects Nos. 331342, and 358088,
InstituteQ, the Jane and Aatos Erkko Foundation, and the Finnish Quantum Flagship.
We acknowledge the computational resources provided by the Aalto Science-IT project.
We thank Tiago Ant\~{a}o, Adolfo Fumega, Rouven Koch, 
Peter Liljeroth and Robert Drost for useful discussions.

\appendix
\section{Neural Network Training}
\label{sec:neural_network}
\subsection{Fermionic Chain}

For the fermionic chain, we employed a fully connected NN architecture. The physical observable used as input to train the network is the spectral function, as defined in Section~\ref{subsec:fermionic_chain}. We computed the spectral function using a Green's function formalism~\cite{pyqula}, where the energy smearing is taken as \( \Delta = 1 \times 10^{-2} t \). We generated a dataset of 5000 samples, with 3000 samples designated for training and 1000 samples for testing. The training process spanned 200 epochs, utilizing a batch size of 32. The Adam optimizer was used for optimization, with a learning rate set to $10^{-4}$. The detailed layer structure of the NN is summarized in Table~\ref{tab:nn_fermions}.

\begin{table}[h]
    \centering
    \begin{tabular}{c|c|c}
        \hline
        \textbf{Layer} & \textbf{Neurons} & \textbf{Activation} \\
        \hline
        Input  & 900  & - \\
        Dense  & 512  & ReLU \\
        Dense  & 512  & ReLU \\
        Dense  & 256  & ReLU \\
        Dense  & 256  & ReLU \\
        Dense  & 128  & ReLU \\
        Dense  & 128  & ReLU \\
        Dense  & 64   & ReLU \\
        Dense  & 64   & ReLU \\
        Dense  & 32   & ReLU \\
        Dense  & 32   & ReLU \\
        Dense  & 16   & ReLU \\
        Dense  & 16   & ReLU \\
        Output & 3    & Linear \\
        \hline
    \end{tabular}
    \caption{Architecture of the NN used for learning the parameters of the fermionic system.}
    \label{tab:nn_fermions}
\end{table}

The input data for the NN was prepared by considering a chain of 150 sites. We evaluated the spectral function for three different voltage biases, $V \in [-0.5, 0, 0.5]$, and then applied a Fourier transform to the spectral function data. This preprocessing step resulted in an input size of 900 for the first layer of the NN. The loss function used for training was the mean squared error (MSE). The chosen parameter ranges are as follows: $\mu$ and $\lambda_R$ ranged from $[0,1]$, while $t_2$ spanned $[0,2]$.

The training process followed a two-step approach. Initially, the model was trained solely on noiseless data, employing an early stopping criterion based on validation loss within the first 5 epochs. After this phase, magnetic noise was added to the original data set at levels of $\Delta^2_B=0.05$ and $\Delta^2_B=0.1$. We then resumed training the same NN for an additional 200 epochs. This methodology allows us to assess the robustness of the model under varying noise levels. At the end of the process, we obtained three models: one trained exclusively on noiseless data and two additional models fine-tuned with increasing levels of noise. The workflow of our NN training for the fermionic chain is illustrated in Fig.~\ref{fig:pic4}a.

\subsection{Spin Chain}

For the spin chain system, we adopted the same fully connected NN architecture, utilizing the static spin correlator as input, as defined in Section~\ref{subsec:spin_chain}. The dataset consisted of 2000 samples, with 1200 samples for training and 400 samples for testing. The training process lasted 100 epochs, using the Adam optimizer with a learning rate of $10^{-4}$ and batch size of 32. The detailed layer structure of the NN is summarized in Table~\ref{tab:nn_spins}.

\begin{table}[h]
    \centering
    \begin{tabular}{c|c|c}
        \hline
        \textbf{Layer} & \textbf{Neurons} & \textbf{Activation} \\
        \hline
        Input  & 20  & - \\
        Dense  & 32  & ReLU \\
        Dense  & 32  & ReLU \\
        Dense  & 16  & ReLU \\
        Dense  & 16  & ReLU \\
        Dense  & 8   & ReLU \\
        Dense  & 8   & ReLU \\
        Output & 3   & Linear \\
        \hline
    \end{tabular}
    \caption{Architecture of the NN used for learning the parameters of the spin chain system.}
    \label{tab:nn_spins}
\end{table}

The input data was derived from a spin chain of 21 sites. We evaluated the static spin correlator for 12 different external magnetic fields in the range $B \in [0, 5]$, resulting in an initial input size of 252. PCA was then applied to reduce the input size to 20 components. The loss function used for training was MSE, with the parameter ranges for $J_2$ and $J_z$ spanning $[-0.5,0.5]$, and $J_{DM}$ spanning $[0,1]$.

We adopted the same two-step training strategy as in the fermionic chain case. Initially, the model was trained on noiseless data, with early stopping applied within the first 5 epochs based on validation loss. Subsequently, magnetic noise at levels of $\Delta^2_B=0.05$ and $\Delta^2_B=0.1$ was added to the dataset. PCA was reapplied, and training resumed for an additional 100 epochs. This process allowed us to systematically evaluate the model's robustness against noise. The workflow of the NN training for the spin chain is illustrated in Fig.~\ref{fig:pic5}a.

We also explored an alternative network architecture with separate branches to learn \( J_2 \), \( J_z \), and \( J_{DM} \). However, this approach did not produce significant performance improvements. Therefore, we adopt a simultaneous learning strategy that ensures consistency across all parameters while maintaining comparable or superior performance.

\bibliography{main}
\end{document}